\documentclass[pra,aps,twocolumn,showpacs,ams]{revtex4}
\usepackage{graphicx}
\usepackage{dcolumn}
\usepackage{latexsym}
\usepackage{color}

\begin{document}



\title{
Pauli-paramagnetic effects on  
mixed state properties in a strongly anisotropic 
superconductor ---Application to Sr$_2$RuO$_4$---
}



\author{Yuujirou Amano} 
\affiliation{
Department of Physics, Okayama University, Okayama 700-8530, JAPAN}

\author{Masahiro Ishihara} 
\affiliation{
Department of Physics, Okayama University, Okayama 700-8530, JAPAN}

\author{Masanori Ichioka}
\email[]{ichioka@cc.okayama-u.ac.jp}
\affiliation{
Department of Physics, Okayama University, Okayama 700-8530, JAPAN}

\author{Noriyuki Nakai} 
\affiliation{
Department of Physics, Okayama University, Okayama 700-8530, JAPAN}

\author{Kazushige Machida} 
\email[]{machida@mp.okayama-u.ac.jp}
\affiliation{
 Department of Physics, Okayama University, Okayama 700-8530, JAPAN}


\date{\today}

\begin{abstract}
We study theoretically 
the mixed state properties of a strong uniaxially-anisotropic type II
superconductor with the Pauli paramagnetic effect, focusing on their behaviors 
when the magnetic field orientation is tilted 
from the conduction layer $ab$ plane.
On the basis of Eilenberger theory,  
we quantitatively estimate significant contributions of 
the Pauli paramagnetic effects on
a variety of physical observables, including  
transverse and longitudinal components of the flux line lattice form factors, 
magnetization curves, Sommerfeld coefficient,
field distributions and magnetic torques. 
We apply these studies to ${\rm Sr_2RuO_4}$ 
and quantitatively explain several seemingly curious behaviors, 
including the $H_{\rm c2}$ suppression for the $ab$ plane direction,  
the larger anisotropy ratio and intensity found by 
the spin-flip small angle neutron scattering,  
and the first order transition observed recently 
in magneto-caloric, specific heat and magnetization measurements
in a coherent and consistent manner.  
Those lead us to conclude that 
${\rm Sr_2RuO_4}$ is either a spin-singlet or a spin-triplet pairing 
with the $d$-vector components in the $ab$ plane. 
\end{abstract}



\maketitle

\section{Introduction}
\label{sec:introduction}

Sr$_2$RuO$_4$ is well known to be a prime candidate of a chiral 
$p$-wave superconductor~\cite{MackenzieMaeno,maeno,bergemann}. 
The crystal structure is same as  in La$_2$CuO$_4$: 
a mother compound of high $T_{\rm c}$
superconductors. The normal state properties of Sr$_2$RuO$_4$
are characterized by quite a standard Landau Fermi liquid picture with
a moderate mass renormalization~\cite{MackenzieMaeno} 
in stark contrast with the high $T_{\rm c}$ cuprates 
which are strange metals in every respect.
Yet both have a strong two dimensional metallic conduction associated
with  anisotropic layered structure. 
In this sense Sr$_2$RuO$_4$ has a firm foundation,
out of which the superconducting state develops at $T_{\rm c}$=1.5K.
Thus we can safely employ a reliable theoretical framework such as Eilenberger
theory that assumes a normal Fermi liquid for 
describing the superconducting properties under
an applied field.

Recently, the research front of Sr$_2$RuO$_4$ has been greatly advanced: 
(1) The small angle neutron scattering (SANS) experiment~\cite{Rastovski} 
shows that the anisotropy ratio of the vortex lattice
amounts to $\Gamma_{\rm VL}\sim60$ 
for the field orientation $\bar{\bf B}$ parallel to the $ab$ plane. 
This is at odds with the $H_{\rm c2}$ anisotropy ratio 
$\Gamma_{H_{\rm c2}}\equiv H_{\rm c2, ab}/H_{\rm c2, c}=20$, 
where $H_{\rm c2, ab}$  ($H_{\rm c2, c}$) is 
the upper critical field $H_{\rm c2}$ for $\bar{\bf B} \parallel ab$ 
($\bar{\bf B} \parallel c$),  
because in usual single-band superconductors 
$\Gamma_{\rm VL} \sim \Gamma_{H_{\rm c2}}$ is 
expected~\cite{MiranovicMachidaKogan}.

(2) The magneto-caloric effect~\cite{kajikawa}, 
the specific heat~\cite{yonezawa} 
and magnetization experiments~\cite{kittaka} detect 
the first order transition at $H_{\rm c2, ab}$ in low temperatures, 
which is similar to superconductors 
with strong Pauli paramagnetic effect (PPE), 
such as in CeCoIn$_5$~\cite{tayama,kaku,kumagai,bianchi}.

We note that the three experiments\cite{kajikawa,yonezawa,kittaka} 
mentioned above are mutually quite consistent with each others,
since a certain amount of quasi-particles 
in the superconducting state
are responsible for exhibiting the first order transition's jumps 
at $H_{\rm c2,ab}$ in those thermodynamic quantities.  
There the same quasi-particles manifest themselves in each observable.
This means that viewing from the normal side above $H_{\rm c2,ab}$
the spin susceptibility $\chi_{\rm spin}$ must decrease 
in the superconducting state. 
However, this expectation is in conflict with the  
existing Knight shift experiments by NMR~\cite{ishida,murakawa,ishida2} 
and polarized neutron scattering measurements~\cite{hayden}.
There are no triplet pairing theories proposed so far 
which are able to predict the first order $H_{{\rm c2},ab}$ transition, 
including works by one of the present 
authors~\cite{rice-sigrist, zhito,annett,nomura,koikegami,yanase,thalmeir,
kuwabara,sato,kuroki,takimoto,dahm,wu,eremin,miyake,hasegawa,kubo,won,graf,choi,klemm,machida-triplet}. 

There are some other experimental reports suggesting the spin-triplet 
chiral $p$-wave superconductivity in ${\rm Sr_2RuO_4}$. 
For example, 
the observation of half-quantized fluxoids~\cite{hqv}, which requires
multiple order parameter for 
the pairing function with both spin and orbital degrees of freedom active,
implying the possibility of the spin-triplet pairing. 
The chiral domain formations
and the time reversal symmetry breaking are suggested 
by various experimental methods~\cite{luke,kid,kashiwaya}. 
However, a scanning Hall probe experiment~\cite{kirtley} 
fails to detect the edge current expected for the chiral superconductors. 
In the experiments, the estimated domain sizes for each sample
used in those experiments are strangely widely different from
1$\mu$m to 1mm (see Ref. \onlinecite{kallin} 
for detailed critical examinations on this point).
Therefore, in the present status of understanding the mechanism of 
superconductivity in ${\rm Sr_2RuO_4}$, the above-mentioned 
experimental results are mutually contradicted.  

The purpose of this paper is to find a clue to resolve the contradictions, 
by describing the mixed state properties
for a uniaxial strong anisotropic type II superconductor with 
PPE in the clean limit and a single band on the basis of 
quasi-classical Eilenberger theory.
Then we critically examine several experiments done recently on Sr$_2$RuO$_4$ 
and interpret the implications of those experiments in the viewpoint of 
the PPE. 
It is shown that the results are maximally consistent 
with the experimental data, and stimulate future 
theoretical and experimental studies to further understand the mechanism 
of the exotic superconductivity. 

The Eilenberger theory is applicable for superconductors with 
$k_{\rm F} \xi \gg 1$.
For Sr$_2$RuO$_4$ this condition is well satisfied because
the coherence length $\xi\sim 30$nm and the inverse of the Fermi wave number
$k_F^{-1}\sim$ a few nm.
In this paper we employ spin-singlet isotropic $s$-wave pairing 
for simplicity to grasp the essential features of the PPE. 
Among the orbital and spin parts of the pairing symmetry, 
essential assumption in the present theory is that 
the PPE works in the spin part of the pairing. 
In addition to the case of the spin-singlet pairing, 
we can expect similar behavior of PPE 
also in the spin-triplet pairing case  
if the $d$-vector has components in the $ab$ plane. 
The assumption for the orbital part as isotropic $s$-wave 
is not intrinsic condition. 
Both the $s$-wave and the $d$-wave pairing show similar high field 
behaviors of PPE~\cite{IchiokaPara}. 
Also in the chiral $p$-wave pairing, 
we see the similar transverse components of the internal 
fields~\cite{ishihara}. 
Thus, the replacement of the orbital part from the $s$-wave pairing 
to the chiral $p$-wave pairing is possible, 
and we expect similar behaviors there,  
if the PPE is active in the spin-part of the pairing.  

The arrangement of this paper is as follows.
In Sec. \ref{sec:formulation} 
we give the formulation based on Eilenberger theory with PPE. 
The spatial structures of vortices, 
including internal magnetic field ${\bf B}({\bf r})$ and 
the paramagnetic moment $M_{\rm para}({\bf r})$, 
are described in Sec. \ref{sec:spatial}.
The form factors responsible for SANS experiments are evaluated  both 
for the longitudinal and transverse components relative to 
the applied field orientation in Sec. \ref{sec:FF}. 
In the next section \ref{sec:FO} we calculate the magnetization curves and 
Sommerfeld coefficient $\gamma(\bar{B})$ as a function of magnetic fields 
to examine the first order transition's jumps 
of these quantities at  $H_{\rm c2}$.
The distributions of $P(B)$ of $B({\bf r})$ and 
$P(M)$ of $M_{\rm para}({\bf r})$ of the vortex lattice state,
that are responsible for the resonance line shape of the NMR spectra,  
are calculated in Sec. \ref{sec:PB-PM}.
The magnetic torque curves are also evaluated in Sec. \ref{sec:torque}.
We discuss 
intrinsic anisotropy of $\Gamma_{\rm VL}$ and $\Gamma_{H_{\rm c2}}$ 
in Sec. \ref{sec:discussion}.
The final section \ref{sec:conclusion} 
is devoted to conclusion and future problems.
The present paper belongs to our series of papers on 
the magnetic field orientation dependence of uniaxial 
superconductors: chiral $p$-wave case~\cite{ishihara} and
s-wave and $d$-wave cases without PPE~\cite{amano}.

\section{Quasiclassical theory including Pauli paramagnetic effect}
\label{sec:formulation}

First, we explain the coordinate and 
the Fermi surface used in our calculations.
We consider the case when the magnetic field orientation 
is tilted by $\theta$ from the $c$ axis towards the $ab$ plane. 
We write the crystal coordinate as $(a,b,c)$.  
To describe the vortex structure,  
we use the coordinate ${\bf r}=(x,y,z)$ 
where $z$ axis is set to the vortex line direction. 
Thus, the relation to the vortex coordinate and the crystal coordinate is 
given by 
$(x,y,z)=(a,b \cos\theta + c \sin\theta,c \cos\theta -b \sin\theta)$.  

As a model of the Fermi surface, we use quasi-two dimensional 
Fermi surface with rippled cylinder shape. 
In the crystal coordinate, the Fermi velocity is assumed to be 
${\bf v}=(v_a,v_b,v_c)\propto(\cos\phi,\sin\phi,\tilde{v}_z \sin p_c)$ 
at 
${\bf p}=(p_a,p_b,p_c)\propto(p_{\rm F}\cos\phi, p_{\rm F}\sin\phi,p_c)$
on the Fermi surface~\cite{Hiragi}.  
From the Fermi surface, 
anisotropy ratio of the coherence lengths is estimated as 
\begin{eqnarray}
\Gamma \equiv \xi_{c} / \xi_{b} \sim  
\langle v_c^2 \rangle_{\bf p}^{1/2}/\langle v_b^2 \rangle_{\bf p}^{1/2} 
\sim 1/\tilde{v}_z, 
\label{eq:Gamma-xi}
\end{eqnarray} 
where $\langle \cdots \rangle_{\bf p}$ 
indicates an average over the Fermi surface. 

The spatial structure of quasiparticles in the superconducting state 
is studied by the Eilenberger theory. 
Quasiclassical Green's functions 
$f(\omega_n , {\bf p},{\bf r})$,  
$f^\dagger(\omega_n , {\bf p},{\bf r})$,  and
$g(\omega_n , {\bf p},{\bf r})$ 
are calculated in the vortex lattice states  
by solving Riccati equation, 
which is derived from 
the Eilenberger equation
\begin{eqnarray} && 
\left\{ \omega_n + {\rm i}{\mu} B({\bf r}) 
        + \hat{\bf v} \cdot 
           \left(\nabla + {\rm i}{\bf A}({\bf r}) \right) \right\} f 
= \Delta({\bf r}) g, 
\nonumber \\  && 
\left\{ \omega_n + {\rm i}{\mu} B({\bf r})  
        - \hat{\bf v} \cdot 
           \left(\nabla - {\rm i}{\bf A}({\bf r})  \right) \right\} f^\dagger  
= \Delta^\ast({\bf r}) g ,  
\label{eq:Eil} 
\end{eqnarray} 
in the clean limit with 
\begin{eqnarray} && 
\hat{\bf v} \cdot {\nabla} g 
= \Delta^\ast({\bf r}) f  
- \Delta({\bf r}) f^\dagger,  
\end{eqnarray}  
$g=(1-ff^\dagger)^{1/2}$, ${\rm Re}g >0$, 
and Matsubara frequency 
$\omega_n$~\cite{Hiragi,Klein,Miranovic,IchiokaPara}.     
The paramagnetic parameter ${\mu}=\mu_{\rm B} B_0/\pi k_{\rm B}T_{\rm c}$ 
is proportional to the Maki parameter. 
We calculate the spatial structure of $g$ in a fully self-consist way
without using Pesch's approximation~\cite{Pesch}. 
We consider the case of isotropic $s$-wave pairing, 
because the paramagnetic effect does not seriously 
depend on the pairing function of the orbital part~\cite{IchiokaPara}.  
Normalized Fermi velocity is $\hat{\bf v}={\bf v}/v_{\rm F}$ with 
$v_{\rm F}=\langle {\bf v}^2 \rangle_{{\bf p}}^{1/2}$.
We have scaled length, temperature, magnetic field, 
and energies 
in units of $\xi_0$, $T_c$,
$B_0$, and $\pi k_{\rm B} T_{\rm c}$, respectively, 
where $\xi_0=\hbar v_{{\rm F}}/2\pi k_{\rm B} T_{\rm c}$ and 
$B_0=\phi_0 /2 \pi \xi_0^2$.
$\phi_0$ is the flux quantum. 
The vector potential 
${\bf A}=\frac{1}{2}\bar{{\bf B}}\times{\bf r}+{\bf a}({\bf r})$
is related to the internal field as 
${\bf B}({\bf r})=\nabla\times {\bf A}
 =(B_x({\bf r}),B_y({\bf r}),B_z({\bf r}))$ 
with $\bar{\bf B}=(0,0,\bar{B})$, 
$B_z({\bf r})=\bar{B}+b_z({\bf r})$ and 
$(B_x,B_y,b_z)=\nabla\times {\bf a}$. 
The spatial averages of $B_x$, $B_y$, and $b_z$ are zero. 
$\bar{B}$ is the averaged flux density of the internal field.  

The pairing potential $\Delta({\bf r})$ in the isotropic $s$-wave pairing 
is calculated by the gap equation 
\begin{eqnarray}
\Delta({\bf r})
= g_0N_0 T \sum_{0 \le \omega_n \le \omega_{\rm cut}} 
\left\langle 
    f +{f^\dagger}^\ast 
\right\rangle_{{\bf p}} 
\label{eq:scD}
\end{eqnarray}
where 
$g_0$ is the pairing interaction 
in the low-energy band $|\omega_n|\le\omega_{c}$, and 
$N_0$ is the density of states (DOS) at the Fermi energy in the normal state. 
$g_0$ is defined by the cutoff energy $\omega_{\rm c}$ as  
$(g_0N_0)^{-1} = \ln T+2\,T\sum_{\omega_n>0}^{\omega_{\rm c}}\,\omega_n^{-1}$.
We carry out calculations using the cutoff $\omega_{\rm c}=20 k_{\rm B}T_{\rm c}$. 
Current equation to obtain ${\bf a}$ is given by 
\begin{eqnarray}
\nabla\times  \nabla \times {\bf a}({\bf r}) 
={\bf j}_{\rm s}({\bf r})+\nabla\times {\bf M}_{\rm para}({\bf r})
\label{eq:rotA}
\end{eqnarray} 
with the screening current 
\begin{eqnarray} && 
{\bf j}_{\rm s}({\bf r})
=-\frac{2T}{{{\kappa}}^2}  \sum_{0 \le \omega_n} 
 \left\langle \hat{\bf v} {\rm Im}\{ g \}  
 \right\rangle_{{\bf p}} , 
\label{eq:scH} 
\end{eqnarray} 
and the paramagnetic moment 
\begin{eqnarray} && 
M_{\rm para}({\bf r})
=M_0 \left( 
\frac{B({\bf r})}{\bar{B}} 
- \frac{2T}{{\mu} \bar{B} }  
\sum_{0 \le \omega_n}  \left\langle {\rm Im} \left\{ g \right\} 
 \right\rangle_{{\bf p}}
\right) .
\label{eq:scM} 
\end{eqnarray} 
Here, the normal state paramagnetic moment 
$M_0 = ({{\mu}}/{{\kappa}})^2 \bar{B} $, and 
${\kappa}=B_0/\pi k_{\rm B}T_{\rm c}\sqrt{8\pi N_0}$.   
The Ginzburg-Landau (GL) parameter $\kappa$ 
is the ratio of the penetration depth to coherence length for 
$\bar{\bf B}\parallel c$.

\begin{figure}
\begin{center}
\includegraphics[width=6.5cm]{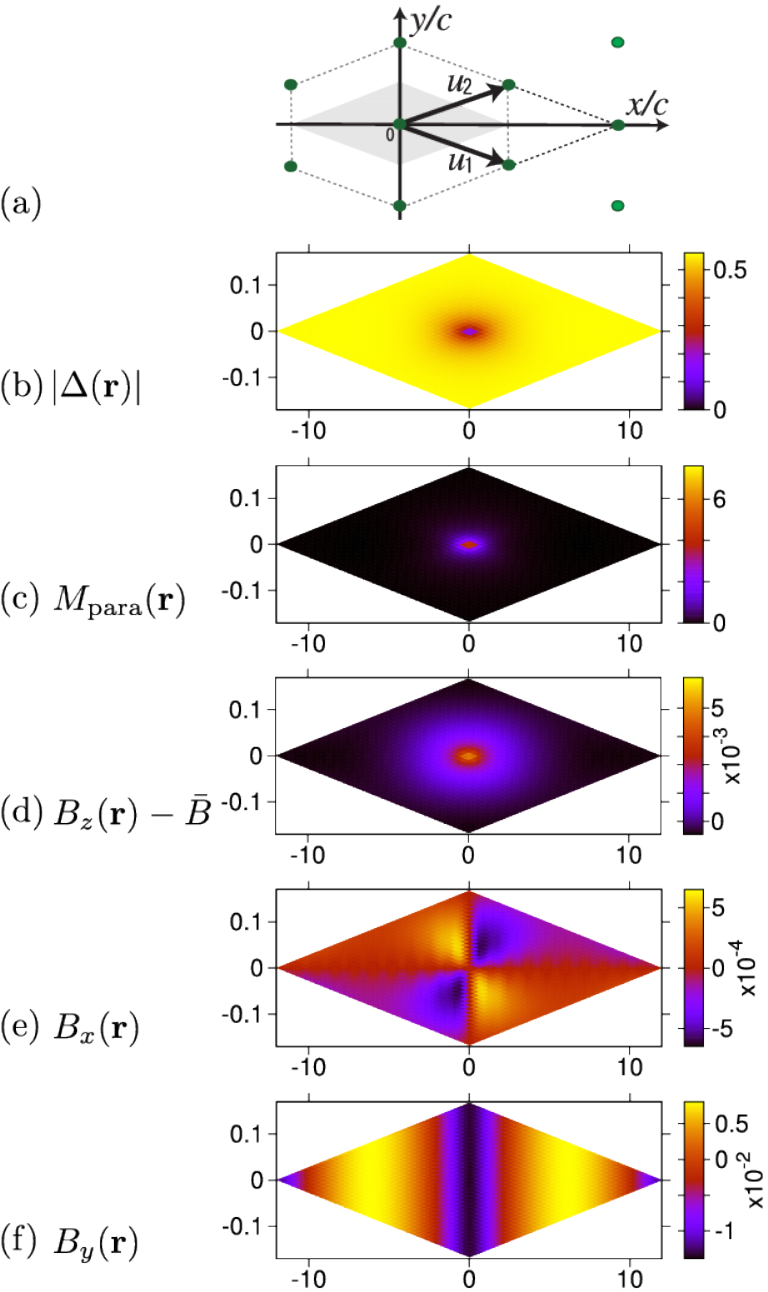}
\end{center}
\caption{\label{fig1}
(Color online) 
(a) Unit vectors ${\bf u}_1$ and ${\bf u}_2$ of the vortex lattice. 
Circles indicate vortex centers. 
Gray region is a unit cell of our calculations. 
(b) $|\Delta({\bf r})|$.
(c) $M_{\rm para}({\bf r})$.  
(d) $B_z({\bf r})-\bar{B}$. 
(e) $B_x({\bf r})$.
(f) $B_y({\bf r})$. 
In (b)-(f), we show density plot within a unit cell, 
when $\theta=89^\circ$ at $\bar{B}=1.5$ and $\mu=0.04$. 
}
\end{figure}

We set unit vectors of the vortex lattice as
\begin{eqnarray} && 
{\bf u}_1=c({\alpha}/{2},-{\sqrt{3}}/{2}), \ 
{\bf u}_2=c({\alpha}/{2}, {\sqrt{3}}/{2})
\label{eq:unit-cell}
\end{eqnarray} 
with $c^2=2 \phi_0/ (\sqrt{3} \alpha \bar{B})$ and 
$\alpha=3 \Gamma(\theta)$~\cite{Hiragi}, 
as shown in Fig. \ref{fig1}(a). 
We use the anisotropic ratio $\Gamma(\theta) \equiv
\xi_{y} / \xi_{x} \sim 
\langle v_y^2 \rangle_{\bf p}^{1/2}/\langle v_x^2 \rangle_{\bf p}^{1/2}$,
that is,
\begin{eqnarray} && 
\Gamma(\theta) 
={1\over{\sqrt{\cos^2\theta+\Gamma^{-2}\sin^2\theta}}} 
\label{eq:Gamma}
\end{eqnarray} 
of the effective mass model. 
Supposing the case of ${\rm Sr_2RuO_4}$~\cite{MackenzieMaeno}, 
we set to be $\kappa=2.7$ and the anisotropy ratio 
$\Gamma(\theta=90^{\circ})=\Gamma=60$, 
which is suggested by the SANS experiment~\cite{Rastovski}.
By the iteration of calculations by Eqs. (\ref{eq:Eil})-(\ref{eq:scM}) 
at $T=0.1T_{\rm c}$, 
we obtain self-consistent solutions of 
$\Delta({\bf r})$, ${\bf A}({\bf r})$, and 
quasiclassical Green's functions. 

From the selfconsistent solutions, 
we calculate the following physical quantities. 
In Eilenberger theory, free energy is given by 
\begin{eqnarray} &&
F=
{\kappa}^2 \langle |{\bf B}({\bf r})-\bar{\bf B}|^2 \rangle_{\bf r}
-{\mu}^2 \langle |{\bf B}({\bf r})|^2 \rangle_{\bf r} 
\nonumber \\ && \qquad
+T  \sum_{|\omega_n|<\omega_{\rm cut}}
\left\langle   {\rm Re} \left\langle 
\frac{g-1}{g+1}(\Delta f^\dagger + \Delta^\ast f )
 \right\rangle_{\bf k} \right\rangle_{\bf r}, 
\label{eq:f3}
\end{eqnarray} 
when Eqs. (\ref{eq:Eil}) and (\ref{eq:scD}) are satisfied~\cite{Hiragi}.  
$\langle \cdots \rangle_{\bf r}$ indicates the spatial average. 
The magnetization is calculated as 
$M_{\rm total}=\bar{B}-H$, where the external field $H$ is given by 
\begin{eqnarray} && 
H=\left(1-\frac{{\mu}^2}{{\kappa}^2}\right) 
\left(\bar{B}
+\left\langle \left( B({\bf r})-\bar{B} \right)^2\right\rangle_{\bf r}
/{\bar{B}} \right)
\nonumber \\ &&   
+\frac{T}{{\kappa}^2 \bar{B}} \langle  \sum_{0 < \omega_n} 
\langle 
{\mu} B_z({\bf r}) {\rm Im} \left\{ g \right\}  
+\frac{1}{2}{\rm Re}\left\{ 
\frac{(f^\dagger \Delta+f \Delta^\ast)g}{g+1} \right\} 
\nonumber \\ &&   
\hspace{1cm}
+\omega_n {\rm Re}\{ g-1 \} 
\rangle_{\bf k}\rangle_{\bf r}, 
\end{eqnarray} 
from Doria-Gubernatis-Rainer scaling~\cite{WatanabeKita,Doria}.  
The paramagnetic and diamagnetic components of the magnetization are, 
respectively, $M_{\rm para}=\langle M_{\rm para}({\bf r}) \rangle_{\bf r}$ and 
$M_{\rm dia}=M_{\rm total}-M_{\rm para}$. 
As the resonance line shape of the NMR spectrum for the Knight shift, 
we calculate the distribution function 
$P(M)=\langle \delta(M-M_{\rm para}({\bf r}) )\rangle_{\bf r}$
from the spatial structure of $M_{\rm para}({\bf r})$. 
On the other hand, in the case of negligible hyperfine coupling, 
the NMR signal shows ``Redfield pattern'' 
given by the distribution function 
$P(B)=\langle \delta(B-B({\bf r}) ) \rangle_{\bf r}$ 
calculated from the internal field $B({\bf r})$. 

When we calculate the electronic states,
we solve Eq. (\ref{eq:Eil}) with 
$ {\rm i}\omega_n \rightarrow E + {\rm i} \eta$.
The local density of states (LDOS) is given by
$N({\bf r},E)=N_{\uparrow}({\bf r},E)+N_{\downarrow}({\bf r},E)$, where 
\begin{eqnarray}
N_\sigma({\bf r},E)=N_0 \langle {\rm Re }
\{
g( \omega_n +{\rm i} \sigma{\mu} B, {\bf k},{\bf r})
|_{{\rm i}\omega_n \rightarrow E + {\rm i} \eta} \}\rangle_{\bf k}
\end{eqnarray}
with $\sigma=1$ ($-1$) for up (down) spin component. 
We typically use $\eta=0.01$.
The DOS is obtained by the spatial average of the LDOS as 
$N(E)=N_\uparrow (E) +N_\downarrow (E)
 =\langle N({\bf r},E) \rangle_{\bf r}$. 
We consider the $\bar{B}$-dependence of 
the Sommerfeld coefficient of the specific heat given by the  
zero-energy DOS $\gamma(\bar{B})=N(E=0)/N_0$, 
and the paramagnetic susceptibility 
$\chi_{\rm spin}(\bar{B})=\langle M_{\rm para}({\bf r}) \rangle_{\bf r}/M_0$. 
These are normalized by the normal state values.

\section{Spatial structures of vortices}
\label{sec:spatial}

To discuss $\bar{B}$-dependence of the internal field distribution 
${\rm B}({\bf r})=\nabla\times{\bf A}$,  
we consider flux line lattice (FLL) form factors  
${\bf F}({\bf q}_{h,k})=(F_{x(h,k)},F_{y(h,k)},F_{z(h,k)})$,  
which is obtained by Fourier transformation  
of the internal field distribution as  
${\bf B}({\bf r})=\sum_{h,k}{\bf F}({\bf q}_{h,k}) 
\exp({\rm i}{\bf q}_{h,k}\cdot{\bf r})$  
with wave vector ${\bf q}_{h,k}=h{\bf q}_1+k{\bf q}_2$.
$h$ and $k$ are integers. 
Unit vectors in reciprocal space 
are given by ${\bf q}_1=(2\pi/c)(1/\alpha,-1/\sqrt{3})$ 
and ${\bf q}_2=(2\pi/c)(1/\alpha,1/\sqrt{3})$.  
The $z$-component $|F_{z(h,k)}|^2$ from $B_z({\bf r})$ 
gives the intensity of conventional non-spin flip SANS. 
The transverse component, $|F_{\rm tr(h,k)}|^2=|F_{x(h,k)}|^2+|F_{y(h,k)}|^2$,  
is accessible by spin-flip SANS experiments~\cite{Kealey,Rastovski}.  

\begin{figure}
\begin{center}
\includegraphics[width=6.0cm]{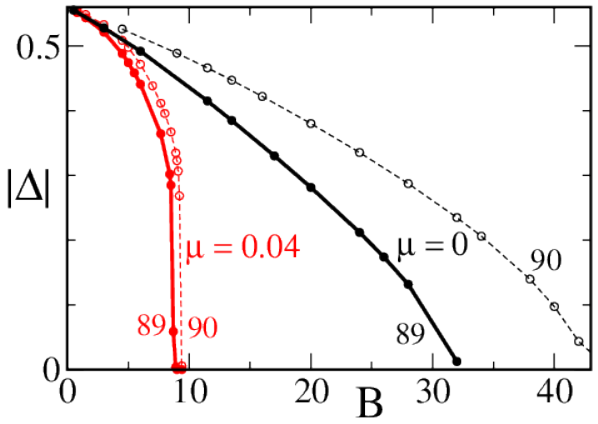}
\end{center}
\caption{\label{fig2}
(Color online) 
(a) $\bar{B}$-dependence of the pair potential 
when $\theta=89^\circ$ (solid lines) and $\theta=90^\circ$ (dashed lines).
Spatial averaged values of 
$|\Delta({\bf r})|$ are presented for $\mu=0$ and $\mu=0.04$. 
The latter ($\mu=0.04$) exhibits first order transitions for both 
$89^\circ$ and $90^\circ$.
}
\end{figure}

First, we study the vortex states 
when the magnetic field orientation is tilted by $1^\circ$ from 
the $ab$ plane ($\theta=89^\circ$). 
In Fig. \ref{fig1}, 
we show the calculated spatial structures 
within a unit cell of vortex lattice at $\bar{B}=1.5$ and $\mu=0.04$. 
The pair potential $\Delta({\bf r})$ has phase winding $2\pi$ 
at the vortex center, and 
the amplitude $|\Delta({\bf r})|$ in Fig. \ref{fig1}(b) is suppressed 
at the vortex core.  
The paramagnetic moment $M_{\rm para}({\bf r})$ in Fig. \ref{fig1}(c) 
is suppressed outside of the vortex core.  
$M_{\rm para}({\bf r})$ appears within the narrow region around the vortex core
and has a large peak at the vortex center. 
As shown in Fig. \ref{fig1}(d), 
the $z$-component of internal field, 
$B_z({\bf r})$, has a peak at the vortex center, 
and decreases as a function of the distance from the center. 
The peak height of $B_z({\bf r})$ is enhanced by the contribution 
of $M_{\rm para}({\bf r})$ at the vortex core~\cite{IchiokaPara}. 
The vortex state has a conventional spatial structure in the  
vortex lattice also when $\bar{\bf B}$ is tilted from the $ab$ plane, 
if length is scaled by the effective coherence length in each direction. 
When $\bar{\bf B}$ is tilted from the $ab$ plane,  
the transverse components $B_x({\bf r})$ and $B_y({\bf r})$ 
appear in the internal field distribution, 
as shown in Figs. \ref{fig1}(e) and \ref{fig1}(f). 
The magnitude of $B_y({\bf r})$ is larger than that of $B_x({\bf r})$. 
The stream lines of $B_y({\bf r})$ in Fig. \ref{fig1}(f) 
flow towards $-y$ direction along vertical stripe region 
connecting vortex cores.
Between the neighbor stripe regions, the stream line flows towards 
$+y$-direction. 
The weak contribution of $B_x({\bf r})$ in Fig. \ref{fig1}(e) 
indicates that the stream lines have weak 
counter-clock-wise (clock-wise) winding 
at positive-$x$ (negative-$x$) region near vortex core. 
These stream line structures of the transverse field is 
qualitatively the same as those obtained by London theory~\cite{Thiemann},    
and as those in a chiral $p$-wave pairing~\cite{ishihara}.

\begin{figure}
\begin{center}
\includegraphics[width=6.0cm]{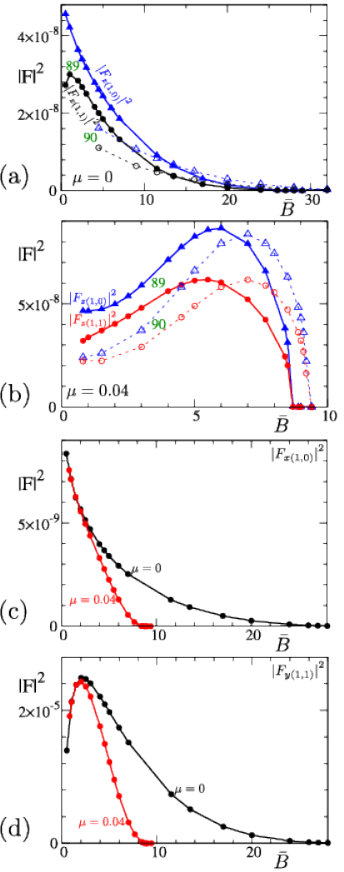}
\end{center}
\caption{\label{fig3}
(Color online) 
$\bar{B}$-dependence of the FLL form factors.  
(a) 
$|F_{z(1,0)}|^2$ and $|F_{z(1,1)}|^2$ for $\mu=0$, 
when $\theta=89^\circ$ (solid lines) and $90^\circ$ (dashed lines). 
(b) The same as (a), but for $\mu=0.04$. 
(c) $|F_{x(1,0)}|^2$ for $\mu=0$ and 0.04 when $\theta=89^\circ$. 
(d) $|F_{y(1,1)}|^2$ for $\mu=0$ and 0.04 when $\theta=89^\circ$. 
}
\end{figure}

When the paramagnetic effect is not considered ($\mu=0$), 
the upper-critical field is $H_{{\rm c2},c}=0.56$ for $\bar{\bf B}\parallel c$
and $H_{{\rm c2},ab}=43$ for $\bar{\bf B}\parallel ab$, 
reflecting large anisotropy $\Gamma$. 
Figure \ref{fig2} presents the amplitude of the pair 
potential as a function of $\bar{B}$. 
In the case $\mu=0.04$, 
the paramagnetic pair-breaking is negligible for $\bar{\bf B}\parallel c$
so that $H_{{\rm c2},c}$ is unchanged. 
However, for  $\bar{\bf B}\parallel ab$, 
the paramagnetic pair breaking becomes eminent at high fields
and limits the upper critical field to $H_{{\rm c2},ab}=9.1$. 
The phase transition at $H_{{\rm c2},ab}$ becomes 
first order as coinciding with the
observation in Sr$_2$RuO$_4$ at low 
temperatures~\cite{kajikawa,yonezawa,kittaka}. 

For the field orientation tilted by $1^\circ$ 
away from the $ab$ plane, namely $\theta=89^\circ$, 
$H_{\rm c2}$ is suppressed from $H_{{\rm c2},ab}$ 
at $\theta=90^\circ$ as seen from Fig. \ref{fig2}.
It is noted that those $H_{\rm c2}$ suppressions are quite different: 
While in the $\mu=0$ case 
$H_{\rm c2}(\theta=89^\circ)/H_{\rm c2}(\theta=90^\circ)=32/43\sim0.74$,
approximately satisfying the expectation based on our effective mass model; 
$\Gamma(\theta=89^\circ)/\Gamma(\theta=90^\circ)\sim 0.69$,
the $H_{\rm c2}$ suppression in the $\mu=0.04$ case is 
very small and remains first order.
This is because $H_{\rm c2}$ is determined 
by the PPE and controlled by the Pauli paramagnetic
critical field $H_p(\theta)$
which has a weak $\theta$ dependence~\cite{IchiokaPara}. 
This point will be discussed later in connection 
with the nature of the phase transitions.

\section{Flux line lattice form factors}
\label{sec:FF}
\subsection{Longitudinal component}

We discuss the $\bar{B}$-dependence of 
the FLL form factor for $\theta=90^\circ$ and $89^\circ$. 
Figures \ref{fig3}(a) and \ref{fig3}(b) present the $\bar{B}$-dependence of 
$|F_{z(1,1)}|^2$ and $|F_{z(1,0)}|^2$. 
These correspond to the intensity of the non-spin-flip SANS experiments. 
When $\mu=0$ in Fig. \ref{fig3}(a), $|F_{z(1,1)}|^2$ and $|F_{z(1,0)}|^2$ 
show exponential decay as a function of $\bar{B}$. 
However, when $\mu=0.04$ in Fig. \ref{fig3}(b), 
both $|F_{z(1,0)}|^2$ and $|F_{z(1,1)}|^2$ increase towards a maximum at 
$\bar{B}\sim 7$ for $\theta=90^\circ$. 
This increasing behavior is due to the enhancement of 
the paramagnetic moment at the vortex, which enhances 
the peak of $B_z({\bf r})$. 
This mechanism\cite{IchiokaPara} was discussed to explain the $\bar{B}$-dependence of 
the SANS intensity in ${\rm CeCoIn_5}$ \cite{bianchi,white} and 
${\rm TmNi_2 B_2C}$~\cite{DeBeer}. 
Compared to the case of $\theta=90^\circ$, 
the intensities of $|F_{z(h,k)}|^2$ are more enhanced 
for $\theta=89^\circ$ at low fields. 
This is because the intensity $|F_{z(h,k)}|^2$ is roughly related to 
the effective GL parameter $\kappa_\theta$ as 
$|F_{z(h,k)}|^2 \propto \kappa_\theta^{-2}$. 
By the anisotropy of $\hat{\bf v}$ in Eq. (\ref{eq:scH}), 
$\kappa_\theta \sim \kappa \Gamma(\theta)$ for the field orientation $\theta$. 
Thus, $\kappa_{89^\circ} < \kappa_{90^\circ}$.  
At high fields,  the peak position of $|F_{z(h,k)}|^2$ is shifted to 
$\bar{B}\sim 6$ when $\theta=89^\circ$,  
reflecting the $\theta$-dependence of anisotropic $H_{\rm c2}$.  

\subsection{Transverse components}

The $\bar{B}$-dependence of the transverse component 
is shown in Figs. \ref{fig3}(c) and \ref{fig3}(d). 
As for $(1,0)$ spot, 
$|F_{{\rm tr}(1,0)}|^2 \sim |F_{x(1,0)}|^2$ since $|F_{y(1,0)}|^2 < 10^{-11}$.  
$|F_{x(1,0)}|^2$ decreases monotonically as a function of $\bar{B}$. 
As for $(1,1)$ spot, 
$|F_{{\rm tr}(1,1)}|^2 \sim |F_{y(1,1)}|^2$ since $F_{x(1,1)} \sim 0$. 
$|F_{y(1,1)}|^2$ decreases as a function of $\bar{B}$, 
after it increases at low $\bar{B}$. 
As in the chiral $p$-wave pairing~\cite{ishihara},  
$|F_{y(1,1)}|^2$ has large intensity, 
compared with $|F_{x(1,0)}|^2$ and $|F_{z(h,k)}|^2$. 
This is consistent to the fact that 
only the spin-flip scattering at $(1,1)$ is observed 
in the SANS experiment\cite{Rastovski} on ${\rm Sr_2RuO_4}$. 
From Figs. \ref{fig3}(c) and \ref{fig3}(d), we see that 
the enhancement due to the paramagnetic effect does not appear 
in the transverse component $|F_{{\rm tr}(h,k)}|^2$. 
Rather $|F_{{\rm tr}(h,k)}|^2$ decreases rapidly at higher fields, 
reflecting the paramagnetic suppression of superconductivity.

\begin{figure}
\begin{center}
\includegraphics[width=6.0cm]{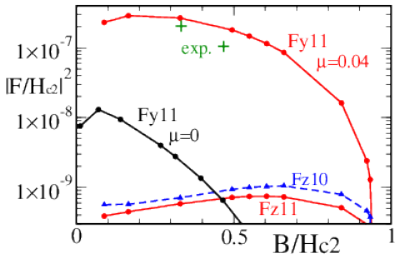}
\end{center}
\caption{\label{fig4}
(Color online) 
$\bar{B}$-dependence of the FLL form factors when $\theta=89^\circ$. 
We plot renormalized values $|F_{y(1,1)}/H_{{\rm c2},ab}|^2$ as a function of 
$\bar{B}/H_{{\rm c2},ab}$ for $\mu=0.04$ and 0. 
The points $+$ indicate experimental values\cite{Rastovski} 
on ${\rm Sr_2RuO_4}$. 
We also present 
$|F_{z(1,0)}/H_{{\rm c2},ab}|^2$ and $|F_{z(1,1)}/H_{{\rm c2},ab}|^2$.  
The vertical axis is a logarithmic scale. 
}
\end{figure}

For the quantitative comparison 
with the experimental data\cite{Rastovski} in ${\rm Sr_2RuO_4}$, 
we discuss the form factors and $\bar{B}$ in unit of 
$H_{{\rm c2}, ab}$ as plotted in Fig. \ref{fig4}.  
In the case $\mu=0.04$,  
$|F_{y(1,1)}/H_{{\rm c2},ab}|^2$ is larger
because $H_{{\rm c2},ab}$ is smaller. 
In Fig. \ref{fig4}, we also show 
the experimental data\cite{Rastovski} 
on ${\rm Sr_2RuO_4}$ with $H_{{\rm c2}, ab}=1.5$[T], 
i.e., 
$F_{{\rm tr}(1,1)}=0.677$[mT] at $\bar{B}=0.5$[T] and 
$F_{{\rm tr}(1,1)}=0.485$[mT] at $\bar{B}=0.7$[T]. 
The magnitude of $|F_{y(1,1)}/H_{{\rm c2},ab}|^2$ in 
experimental data can be quantitatively reproduced in 
the case $\mu=0.04$ including the effect of $H_{{\rm c2},ab}$ suppression. 
From Fig. \ref{fig4}, we also see that $10^{-2}$-times finer resolution 
is necessary in the SANS experiment to observe the spot of $|F_{z(h,k)}|^2$ 
for the non-spin-flip scattering, which is expected to be 
an increasing function of $\bar{B}$ at the middle field range.  

\begin{figure}
\begin{center}
\includegraphics[width=6.0cm]{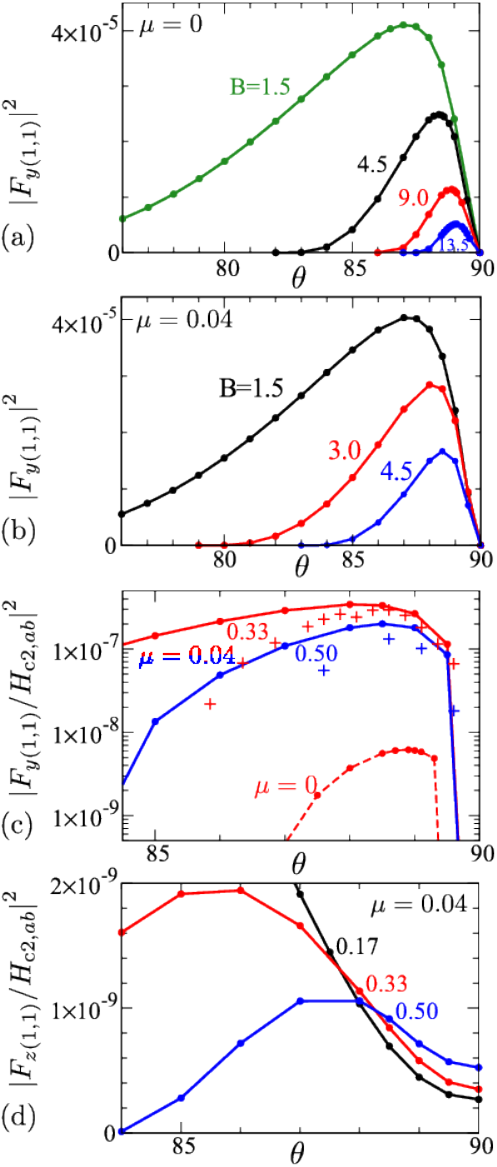}
\end{center}
\caption{
\label{fig5}
(Color online) 
Field orientation $\theta$-dependence of the transverse FLL form factor. 
(a) $|F_{y(1,1)}/H_{{\rm c2},ab}|^2$ as a function of $\theta$ for $\mu=0$ 
    at $\bar{B}=1.5$, 4.5, 9.0, and 13.5. 
(b) $|F_{y(1,1)}/H_{{\rm c2},ab}|^2$ as a function of $\theta$ for $\mu=0.04$ 
    at $\bar{B}=1.5$, 3.0, and 4.5. 
(c) $|F_{y(1,1)}/H_{{\rm c2},ab}|^2$ in a logarithmic scale 
    as a function of $\theta$ for $\mu=0.04$ at 
    $\bar{B}/H_{{\rm c2},ab}\sim0.33$ and 0.5 ($\bar{B}=3.0$ and 4.5). 
    The points $+$ indicate experimental values\cite{Rastovski} 
    on ${\rm Sr_2RuO_4}$ 
    at 0.5[T] and 0.7[T]. 
    We also plot $|F_{y(1,1)}/H_{{\rm c2},ab}|^2$ for $\mu=0$ 
    at $\bar{B}=9.0$ ($\bar{B}/H_{{\rm c2},ab} =0.21$). 
(d) $\theta$ dependence of the longitudinal FLL form factor 
    $|F_{z(1,1)}/H_{{\rm c2},ab}|^2$ for $\mu=0.04$ 
    at $\bar{B}/H_{{\rm c2},ab}=0.17$, 0.33, and 0.50. 
}
\end{figure}

The $\theta$-dependence of the $|F_{y(1,1)}|^2$ is 
presented in Fig. \ref{fig5}. 
As a function of $\theta$,
$|F_{y(1,1)}|^2$ increases until a peak near $90^\circ$. 
After the peak it decreases rapidly towards zero just at $90^\circ$. 
At low enough  field $\bar{B}=1.5$, 
$|F_{y(1,1)}|^2$ shows similar behavior both for $\mu=0$ and 0.04. 
With increasing $\bar{B}$, the peak position is shifted to higher $\theta$, 
and the amplitude is decreased. 
At higher fields such as $\bar{B}=13.5$, 
$|F_{y(1,1)}|^2$ becomes very small for $\mu=0$. 
However, these high field regions vanish for $\mu=0.04$
because of high-field suppression of superconductivity. 
For the quantitative comparison, 
Fig. \ref{fig5}(c) shows renormalized values 
$|F_{y(1,1)}/H_{{\rm c2},ab}|^2$ in a logarithmic scale 
with the SANS results on ${\rm Sr_2RuO_4}$ 
for two cases $\bar{B}/H_{{\rm c2},ab} \sim 0.33$ ($\bar{B}=3.0B_0$ and 0.5[T]) 
and $\bar{B}/H_{{\rm c2},ab} \sim 0.5$ ($\bar{B}=4.5B_0$ and 0.7[T]). 
We see that the experimental data are well fit 
by the theory for $\mu=0.04$ near $\theta=89^\circ$. 
The theoretical values for $\mu=0$ is very small compared to the SANS results. 

We also plot the longitudinal component 
$|F_{z(1,1)}/H_{{\rm c2},ab}|^2$ for $\mu=0.04$ in Fig. \ref{fig5}(d) 
that is not yet observed in ${\rm Sr_2RuO_4}$.
It is seen that  $|F_{z(1,1)}/H_{{\rm c2},ab}|^2$ grows as $\bar{B}$ increases at
$\theta=90^\circ$ because of PPE, as seen in Fig. \ref{fig3}(b).  
At a low field $\bar{B}/H_{{\rm c2},ab}$= 0.17, 
$|F_{z(1,1)}|^2$ monotonically increases when $\theta$ decreases, 
since $|F_{z(1,1)}|^2 \propto \kappa_\theta^{-2}$.  
We notice that the longitudinal components
of the form factor are already observed for $\theta=0$, i.e., 
$H\parallel c$~\cite{forgan2}.
The detailed analysis of those form factors has not done yet, but it seems to
be similar to the results for the square lattice for 
the $d$-wave case~\cite{ichioka}. 
For larger fields $\bar{B}/H_{{\rm c2},ab}$= 0.33 and 0.50, 
it takes a peak at finite $\theta$
because the effective magnetic field $\bar{B}/H_{\rm c2}$ becomes large 
as $\theta$ decreases from  $90^\circ$.
Then $F_{z(1,1)}$ vanishes ultimately towards $H_{\rm c2}$
where the order parameter is zero. 
This peak behavior in $F_{z(1,1)}$ is similar 
to those shown in Fig. \ref{fig3}(b)
where the longitudinal components as a function of  $\bar{B}$ exhibit peaks 
just below  $H_{\rm c2}$.

\section{Jumps at first order $H_{\rm c2}$ transition}
\label{sec:FO}
\subsection{Magnetization curves}

We calculate the magnetization curves for $\theta=89^\circ$
and $90^\circ$ both in the cases of $\mu=0$ and $\mu=0.04$ 
at $T=0.1T_{\rm c}$ as 
shown in Figs. \ref{fig6}(a) and \ref{fig6}(b). 
In the $\mu=0$ case, $M_{\rm total}(\bar{B})$ corresponds to that 
of an ordinary type II superconductor, 
because $M_{\rm total}(\bar{B})$ comes exclusively from the 
orbital diamagnetism due to the orbital current. 
Since the second order transition occurs at $H_{\rm c2}$ in this case, 
$M_{\rm total}(\bar{B})$ smoothly becomes zero.

As seen from Fig. \ref{fig6}(b) in $\mu=0.04$, 
$M_{\rm total}(\bar{B})$ exhibits the jumps $\Delta M_{\rm total}$
at $H_{\rm c2}$ both for $\theta=89^\circ$ and $90^\circ$, 
corresponding to the first order transition. 
The magnetization jump $\Delta M_{\rm total}$ consists of the two components; 
the orbital diamagnetism $\Delta M_{\rm dia}$ and 
the spin paramagnetism $\Delta M_{\rm para}$.
For $\theta=90^\circ$ $M_{\rm para}=1.3\times10^{-3}$ and 
$M_{\rm dia}=-0.7\times10^{-4}$ at $\bar{B}=H_{\rm c2}=9.1$.
Thus $\Delta M_{\rm total}$ is dominated by the spin paramagnetic component.
Since at $\bar{B}=H_{\rm c2}$, $M_{\rm normal}=2.0\times10^{-3}$, 
the relative jump $\Delta M_{\rm total}/M_{\rm normal}=38.5\%$.
As seen from Fig. \ref{fig6}(b), the $\theta=89^\circ$ case also 
gives rise to a similar $\Delta M_{\rm total}$ value.

This number is favorably compared with the experimental value 
$\Delta M_{\rm total}/M_{\rm normal}=15\%$ in Sr$_2$RuO$_4$
at low temperatures~\cite{kittaka}. 
A slightly larger value of 
$\Delta M_{\rm total}/M_{\rm normal}$ in our calculation can be
remedied by considering the multiband effect 
because near $H_{\rm c2}$ the minor band
may be almost in the normal state where the minor gap is already vanishing. 
The minor band contribution can be estimated 
as $\Delta M_{\rm total}/(M_{\rm normal}+M_{\rm minor})\sim19\%$
because the DOS of the minor band is a half of the total DOS.

As seen from Fig. \ref{fig6}(c), 
the contribution of the orbital diamagnetism $M_{\rm dia}$
to the first order jump amounts to $\Delta M_{\rm dia}=-0.7\times 10^{-4}$. 
The weight of the jump, $\Delta M_{\rm dia}/M_{\rm normal}=3.5\%$, 
is an order too smaller than the observed value. 
Thus without PPE it is impossible to understand the large magnetization jump.
We also point out that the magnetization curve for the chiral $p$-wave 
case (see Fig. \ref{fig6}(a) 
in Ref. \onlinecite{ishihara}) is almost same as in usual type II
superconductor without PPE shown in Fig. \ref{fig6}(a).
Thus, if we assume a hypothetical first order transition at $H_{\rm 1st}$,
then $H_{\rm 1st}\sim 0.25H_{\rm c2}$ to account for the magnetization jump 
$\Delta M_{\rm total}/M_{\rm normal}=15\%$.
So far there is no known theory to explain the first order transition 
in the chiral $p$-wave pairing.

\subsection{Specific heat jump at $H_{\rm c2}$}

\begin{figure}
\begin{center}
\includegraphics[width=6.0cm]{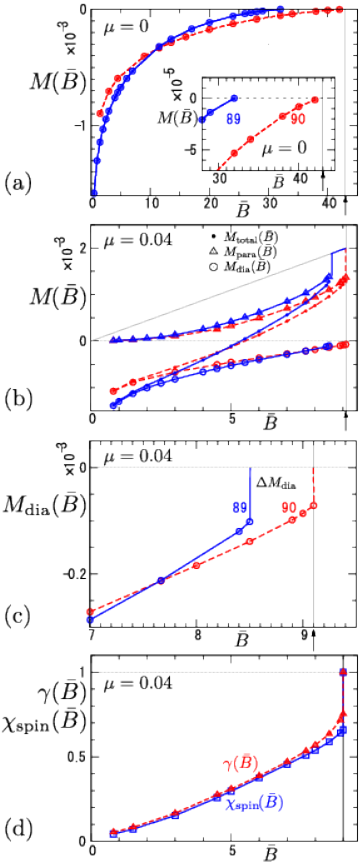}
\end{center}
\caption{\label{fig6}
(Color online) 
$\bar{B}$-dependence of the magnetization.  
(a) $M_{\rm total}(\bar{B})=M_{\rm dia}(\bar{B})$ for $\mu=0$, 
when $\theta=89^\circ$ (solid line) and $90^\circ$ (dashed line). 
In the inset $M_{\rm total}(\bar{B})$ is enlarged near $H_{\rm c2}$.
(b) $M_{\rm total}(\bar{B})$, $M_{\rm para}(\bar{B})$ and $M_{\rm dia}(\bar{B})$ 
 for $\mu=0.04$ when $\theta=89^\circ$ (blue solid lines) and 
$90^\circ$ (red dashed lines).  
(c) $M_{\rm dia}(\bar{B})$ is focused near $H_{\rm c2}$ 
for $\theta=89^\circ$ and $90^\circ$ 
to see the jumps of $\Delta M_{\rm dia}$ at $H_{\rm c2}$.  
(d) The scaling behaviors of $\gamma (\bar{B})$ and $\chi_{\rm spin}(\bar{B})$ 
for $\theta=90^\circ$.
The jumps of $\Delta\gamma$ and $\Delta\chi_{\rm spin}$  
relative to its normal values are seen 
at the first order $H_{\rm c2}$ transition.
}
\end{figure}

We show the calculated results of the DOS at the Fermi level 
at low temperature $T=0.1T_{\rm c}$ in Fig. \ref{fig6}(d), 
which corresponds to the Sommerfeld coefficient $\gamma(\bar{B})$, namely $C/T$
at low temperatures in the superconducting state.
It is known by the explicit calculations~\cite{IchiokaPara} that 
$\gamma(\bar{B})$ is approximately scaled to the
spin susceptibility $\chi_{\rm spin}(\bar{B})$ 
as is seen from Fig. \ref{fig6}(d).
This is because both quantities $\chi_{\rm spin}(\bar{B})$ and 
$\gamma(\bar{B})$ come from the
same DOS of the quasi-particles near the Fermi level~\cite{IchiokaPara}. 
In fact the experimental value\cite{yonezawa} of 
the specific heat jump at $H_{\rm c2}$ is
$\Delta\gamma/\gamma_{\rm normal}\sim10\%$, roughly coinciding with 
$\Delta M_{\rm total}/M_{\rm normal}=15\%$ mentioned above. 
As is seen from Fig. \ref{fig6}(d), 
the jump of $\gamma(\bar{B})$ is slightly smaller 
than that of $\chi_{\rm spin}$ because two quantities are not exactly identical
where the former is an integration of the DOS over $\mu_{\rm B}\bar{B}$
while the latter is the DOS at the Fermi level.

It should be noted that the entropy jump\cite{kajikawa} 
probed by the magneto-caloric measurement
is consistent with the specific  heat jump as discussed 
in Ref. \cite{yonezawa}, meaning that
three experiments, magneto-caloric, specific heat and magnetization 
are mutually consistent with each other. 
If this identification is true, the Knight shift should 
decrease as shown in Fig. \ref{fig6}(d), 
which is contrasted with the claim by the NMR 
experiments\cite{ishida,murakawa,ishida2}
where the Knight shift remains unchanged, 
irrespective to nuclear species ($^{17}$O, $^{87}$Sr, $^{101}$Ru, and $^{99}$Ru),
the field orientations and the field values. 
This is quite at odds in the present analysis.

\section{Field distributions}
\label{sec:PB-PM}

\begin{figure}
\begin{center}
\includegraphics[width=6.0cm]{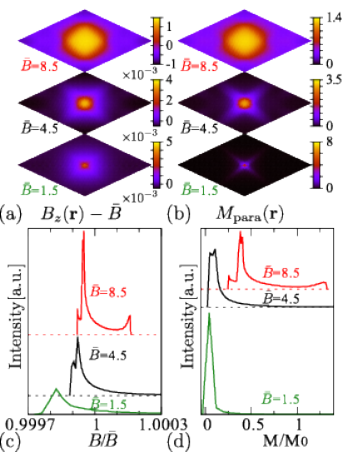}
\end{center}
\caption{\label{fig7}
(Color online)
Topographic maps of (a) $B_z({\bf r})-\bar{B}$ and 
(b) $M_{\rm para}({\bf r})$ within one unit cell 
at $\bar{B}=1.5$, 4.5 and 8.5 for $\mu=0.04$ and $\theta=90^\circ$. 
The field distribution (c) $P(B)$ and (d) $P(M)$ 
associated to (a) and (b), respectively. 
}
\end{figure}

Figures \ref{fig7}(a) and \ref{fig7}(b) display 
the field evolutions of  $P(B)$ and $P(M)$ 
together with the contour maps of 
$B_z({\bf r})$ and $M_{\rm para}({\bf r})$ within a unit cell.
It is seen that as increasing field towards $H_{\rm c2}$ 
the vortex core site and its surrounding sites exclusively 
accommodate the paramagnetic moments induced 
by PPE where the highest $B_z$ and $M_{\rm para}$ are situated.
The mean value of $P(M)$ equals $\chi_{\rm spin}$ in Fig. \ref{fig6}(d). 
At $\bar{B}$=8.5, due to the contributions of 
the paramagnetic moment enhanced at the vortex core, 
$P(B)$ and $P(M)$ have larger weights near the small peak at highest edges, 
whose positions of highest edge exceed $\bar{B}$ and $M_0$, respectively. 
Thus the so-called Redfield pattern $P(B)$  is strongly modified 
from the standard asymmetric distribution in ordinary 
superconductors~\cite{takigawa,tanaka}, such as Nb~\cite{yaouanc}. 
This is also true for $P(M)$ where the asymmetric pattern is modified so that 
the higher $M$ range of the spectrum is enhanced. 

Those asymmetric spectrum patterns should be observed by NMR experiments,  
where neither asymmetric $P(B)$ nor $P(M)$ patterns are not observed 
in any nuclear species ($^{17}$O, $^{87}$Sr, $^{101}$Ru, and $^{99}$Ru ) 
for $\bar{\bf B}\parallel ab$. 
They remain the same patterns as in the normal state~\cite{ishida}. 
Note that the characteristics in Fig. \ref{fig7} are indeed observed 
in CeCoIn$_5$~\cite{kumagai}. 
On the other hand, for $\bar{\bf B}\parallel c$,  
a clear Redfield pattern is observed 
by the muon spin resonance experiment~\cite{forgan1}.  
By analyzing this pattern they correctly reduce the vortex lattice symmetry, 
namely a square lattice that is confirmed later 
by SANS experiments~\cite{forgan2}. 

\section{Magnetic torque}
\label{sec:torque}

\begin{figure}
\begin{center}
\includegraphics[width=6.0cm]{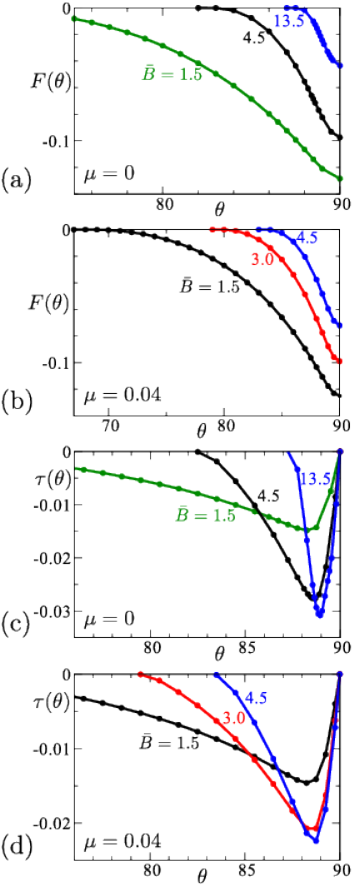}
\end{center}
\caption{\label{fig8}
(Color online)
The $\theta$-dependences of the free energies for various fields $\bar{B}$. 
(a) $\mu=0$ and (b) $\mu=0.04$. 
The corresponding torque curves $\tau(\theta)=dF/d{\theta}$.
(c) $\mu=0$ and (d) $\mu=0.04$.
}
\end{figure}

Since we obtained the self-consistent solutions of 
Eilenberger equation under a given $T$ and $\bar{B}$,
it is not difficult to calculate the magnetic torque 
$\tau(\theta)=dF/d\theta$ by using the free energy $F$
as a function of $\theta$. 
The obtained free energy $F(\theta)$ is displayed 
in Figs. \ref{fig8}(a) and \ref{fig8}(b) for $\mu=0$
and $\mu=0.04$, respectively. 
It is seen that for both cases all the free energy curves smoothly become zero 
when $\theta$ decreases away from $\theta=90^{\circ}$, meaning 
that those are all second order $H_{\rm c2}$ transitions 
in the field range $\bar{B} \le H_{\rm c2}(\theta=88^\circ)$. 

Figures \ref{fig8}(c) and \ref{fig8}(d) show 
the magnetic torque curves $\tau(\theta)$ for $\mu=0$
and $\mu=0.04$, respectively. 
It is seen from those that the sharp minima in $\tau(\theta)$
for both cases are located just near $\theta=90^{\circ}$.  
The fact that the minimum position $\theta_{\rm min}$ in $\tau(\theta)$
is confined near $\theta=90^{\circ}$ is due to 
the large uniaxial anisotropy $\Gamma=60$.
This behavior is easily fit 
by the Kogan torque formula~\cite{kogan} 
based on the London theory: 
\begin{eqnarray} && 
\tau(\theta)\propto 
\frac{\sin2\tilde{\theta}}
     {\sqrt{\cos^2\tilde{\theta}+\Gamma^2\sin^2\tilde{\theta}}}
\ln \frac{\tilde{\eta}\Gamma H_{c2,c}}
     {\bar{B}\sqrt{\cos^2\tilde{\theta}+\Gamma^2\sin^2\tilde{\theta}}}
\nonumber \\ && 
\end{eqnarray} 
with $\tilde{\theta}=90^\circ-\theta$, 
where $\tilde{\eta}$ is a coefficient with the order $\sim 1$.
The minimum $\theta_{\rm min}$ occurs at $\theta_{\rm min}=88.7^{\circ}$ for
$\Gamma=60$ with $\eta=1.5$, which is consistent with our Eilenberger solution.
It should be noticed that at lower fields $\tau(\theta)$ is 
insensitive of the presence or absence of PPE
according to our results in Figs. \ref{fig8}(c) and \ref{fig8}(d). 
Thus, both cases are described by the Kogan formula
which only depends on $\Gamma$.
In fact the minima observed experimentally show 
$\theta^{\rm exp}_{\rm min}\sim89^{\circ}$ at higher fields, 
which becomes $\theta^{\rm exp}_{\rm min}\sim88^{\circ}$ towards lower fields 
(see Fig. 3(d) by Ref. \onlinecite{kittaka}). 
Also in the numerical calculation with $\Gamma=60$, 
$\theta_{\rm min}$ shows similar weak $\bar{B}$-dependence.
Thus the torque data support the large uniaxial anisotropy 
with $\Gamma=60$ for Sr$_2$RuO$_4$.
We note that if we choose $\Gamma=20$ as indicated by $H_{c2,ab}/H_{c2,c}\sim20$,
we find $\theta_{\rm min}\sim87^{\circ}$
which is far off the experimental data\cite{kittaka} 
within the present experimental accuracy.
Thus the intrinsic anisotropy of Sr$_2$RuO$_4$ should be $\Gamma=60$ 
rather than $\Gamma=20$.
The latter number is now understood 
as arising from the suppressed $H_{c2}$ by PPE.

\begin{figure}
\begin{center}
\includegraphics[width=6.0cm]{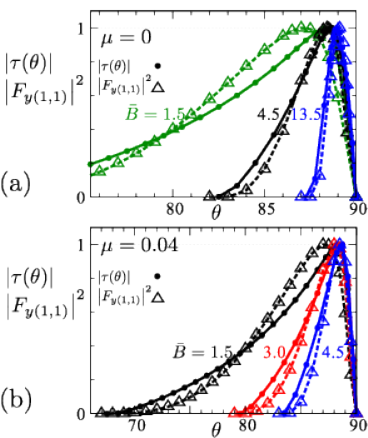}
\end{center}
\caption{\label{fig9}
(Color online) 
The scaling behaviors between $|F_{y (1,1)}|^2$ 
and $|\tau(\theta)|$ as a function of $\theta$. 
(a) $\mu=0$ and (b) $\mu=0.04$. 
Magnitude of each quantity is scaled by the maximum value.  
}
\end{figure}

Since the magnetic torque is related to the transverse components of 
the internal field, it is interesting to compare 
the $|\tau(\theta)|$ curves with the form factor $|F_{y(1,1)}|^2$ 
for both $\mu=0$ and $\mu=0.04$ 
as shown in Figs. \ref{fig9} (a) and \ref{fig9}(b). 
An approximate scaling relationship between them is seen from both cases. 
In particular, 
the maximum position $\theta_{\rm max}$ in both quantities yields the
same value for the higher field data. 
This is indeed seen experimentally 
(see Figs. 3(c) and 3(d) in Ref. \onlinecite{kittaka}).


\begin{figure}
\begin{center}
\includegraphics[width=5.5cm]{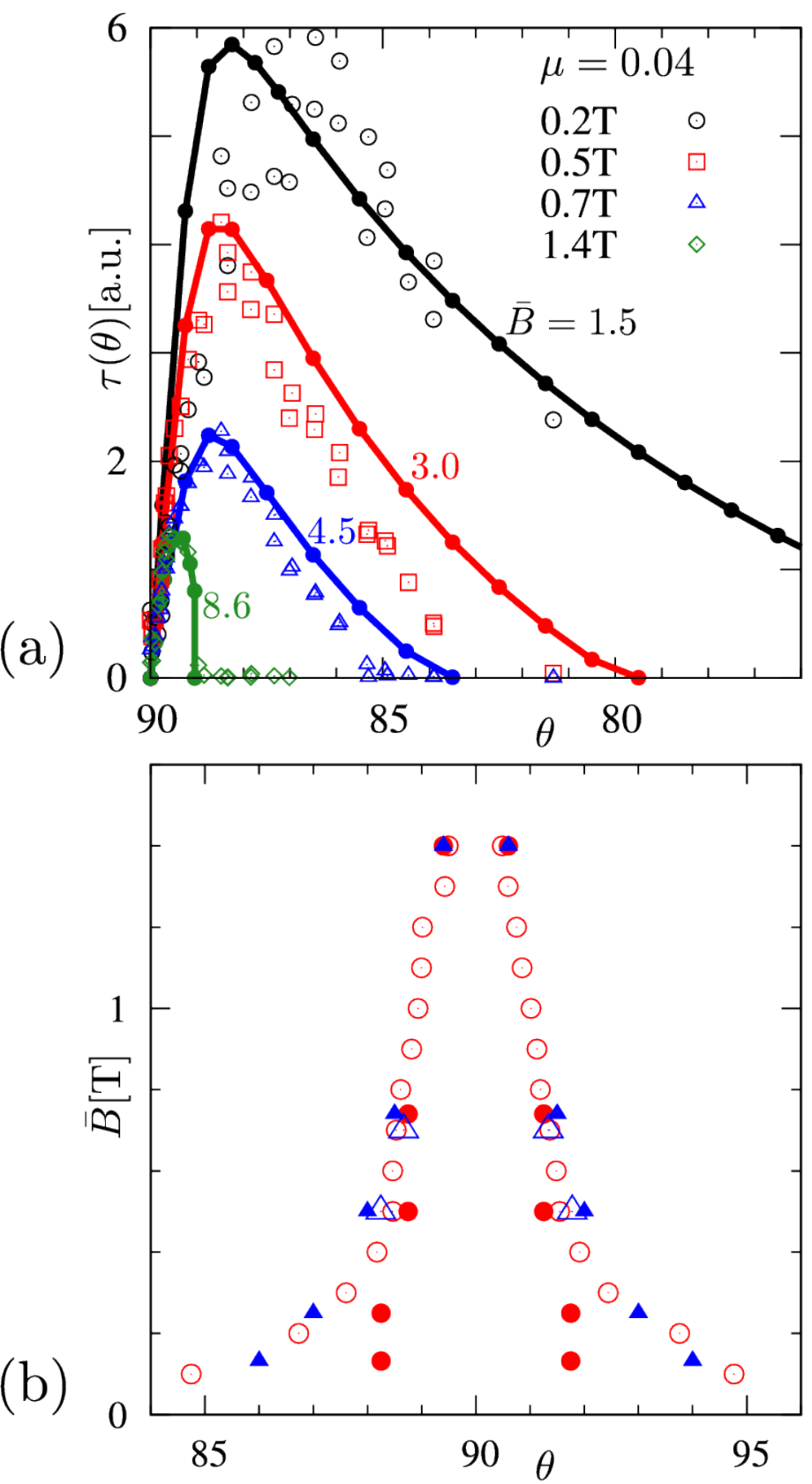}
\end{center}
\caption{\label{fig10}
(Color online) 
(a) Comparison with the theoretical torque curves $|\tau(\theta)|$ 
for $\bar{B}=1.5$, 3.0, 4.5, 8.6 (lines) 
and experimental data~\cite{kittaka} 
for $\bar{B}=0.2$, 0.5, 0.7, 1.4[T] (empty symbols).    
We have adjusted the maximum values of the torque curves 
and displayed those curves by changing its maximum values 
arbitrarily to be clearly seen.
The highest field theoretical data $\bar B=8.6$ and experimental data 1.4[T] 
clearly show the jumps associated with the first order transition.
 (b) 
Maximum angles 
of the form factors (triangles)
and the torque curves (circles) 
in the $\bar B$ and angle $\theta$ plane.  
Theoretical results (filled symbols) are compared with the corresponding 
experimental data~\cite{Rastovski,kittaka} (empty symbols). 
In the scale of vertical axis, $H_{{\rm c2},ab}=9.1$ in theoretical estimate 
is assigned to be 1.5[T].
}
\end{figure}

In Fig. \ref{fig10}(a) 
we compare the theoretical torque curves and corresponding 
experimental data~\cite{kittaka} for selected field values.
It is seen from it that they show a good agreement, 
in particular in the higher field data,
including the maximum angles and vanishing angles of the torque curves.
The highest field theoretical curve $\bar B$=8.6 exhibits 
a first order jump at $\theta=89.1^{\circ}$,
which nicely coincides with the data at 1.4[T].
On the other hand, the lower field data at 0.2[T] show 
a deviation from the theoretical curve $\bar B$=1.5 in their maximum angles. 
This may come from the multi band effect, which will be discussed in the 
forthcoming paper~\cite{nakai}.

In Fig. \ref{fig10}(b) we summarize our maximum angle data coming from
the torque curves and the transverse form factors and compare those
with the experimental data~\cite{Rastovski,kittaka}.
As already indicated in Fig.\ref{fig9}, the discrepancies of the maximum angles
between the torque and form factor occur when the field is lowered.
Since the form factor data at lower fields is lacking at present,
we cannot judge whether or not 
those discrepancies are strengthened further by  future
SANS experiments.
Except for those lowest field data the overall agreement seems 
to be satisfactory.
In other words, the present single band theory gives a reasonable
explanation to those data.

\section{Discussions on phase diagram and intrinsic anisotropy}
\label{sec:discussion}

\begin{figure}
\begin{center}
\includegraphics[width=5.0cm]{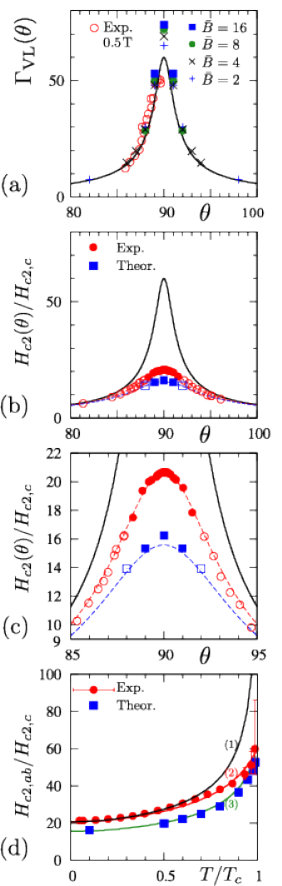}
\end{center}
\caption{\label{fig11}
(Color online) 
(a) The $\theta$-dependences of 
the vortex lattice anisotropy $\Gamma_{\rm VL}(\theta)$.
Open circles indicate the experimental data for $B$=0.5[T]~\cite{Rastovski}. 
Other symbols are for $\Gamma_{\rm VL}(\theta)$ evaluated 
by the free energy minimum at $\bar{B}/B_0=$2, 4, 8, and 16 for $\mu=0$. 
The line presents $\Gamma(\theta)$ of the effective mass model 
in Eq. (\ref{eq:Gamma}) with $\Gamma=60$.
(b) The $\theta$-dependences of $H_{\rm c2}(\theta)/H_{\rm c2,c}$ 
at $T=0.1T_{\rm c}$ for $\mu=0.04$.
(c) Enlarged figure of (b). 
The results of numerical calculations by Eilenberger theory 
are presented by square points. 
The experimental data\cite{kittaka} are shown by circles. 
There, the filled (empty) symbols indicate the first (second) order transition. 
The solid line shows $\Gamma(\theta)$ of the effective mass model 
with $\Gamma=60$.
The  dashed lines correspond to the theoretical curves 
calculated by Eq. (\ref{eq:a3}) with $\Gamma=60$, where 
$\mu=0.04$ for fitting to numerical calculations and 
$\mu=0.0293$ for fitting to experimental data.
(d) 
The anisotropy 
$\Gamma_{H_{\rm c2}}(T)=H_{{\rm c2},ab}(T)/H_{{\rm c2},c}(T)$ as a function of $T$.
The experimental data\cite{kittaka2} (circles) 
and the numerical results by Eilenberger theory with 
$\mu=0.04$ (squares) are shown. 
The three continuous lines are evaluated 
by Eqs. (\ref{eq:a3}) and (\ref{eq:a4}) with 
(1) $\Gamma=180$ and $\mu$=0.0293,
(2) $\Gamma=60$ and $\mu=0.0293$, and
(3) $\Gamma=60$ and $\mu=0.04$.
}
\end{figure}

In previous sections, 
the vortex lattice anisotropy $\Gamma_{\rm VL}(\theta)\equiv\alpha/3$ 
in the definition of Eq. (\ref{eq:unit-cell})  
is assumed to be given by the effective mass model in Eq. (\ref{eq:Gamma}). 
We also perform calculations to determine 
$\Gamma_{\rm VL}(\theta)$ 
by the minimization procedure of the free energy, 
which is much time consuming process compared above. 
The results are shown in Fig. \ref{fig11}(a).
It is seen that this yields a slightly larger $\Gamma_{\rm VL}(\theta)$ 
compared with the effective mass model shown by a line there
around $\theta=90^\circ\pm2^\circ$ region,
beyond which all data points tend to coincide with a line 
of the effective mass model.
We confirm that this deviation of $\Gamma_{\rm VL}(\theta)$ 
does not alter our results in previous sections in a serious way.

We note that, as presented in Fig. \ref{fig11}(a), 
experimental data~\cite{Rastovski} 
also slightly deviate from the effective mass model 
for $87^\circ < \theta < 89^\circ$,  
which is similar to $\Gamma_{\rm VL}(\theta)$ by the free energy minimum. 
This behavior will be discussed in a forth coming paper 
based on multiband model~\cite{nakai}.
From the $\theta$-dependence of $\Gamma_{\rm VL}(\theta)$ 
in Fig. \ref{fig11}(a), 
the intrinsic uniaxial anisotropy of the system can be 
identified as $\Gamma=60$.  
This number just corresponds to the Fermi velocity anisotropy
of the $\beta$ band, since the band-dependent anisotropies are 
estimated as $\Gamma_\alpha=117$, $\Gamma_\beta=57$, and 
$\Gamma_\gamma=174$ for the $\alpha$, $\beta$, and $\gamma$ 
Fermi-surface sheets, respectively, 
according to the dHvA experiments\cite{MackenzieMaeno}.  
We emphasize that this is not accidental, if the $\beta$ band 
plays a major role to govern the mixed state properties 
of the total system in high fields, 
further suggesting that the $\gamma$ band plays a secondary role, 
which is contrary to what many previous works claim, 
such as Nomura and Yamada~\cite{nomura}.

According to the present analysis, 
the $H_{{\rm c2},ab}$ suppression is explained by PPE. 
We evaluate $H_{\rm c2}(\theta)$ for each $\theta$ by the estimate 
of the critical field where  
the order parameters vanish on raising $\bar{B}$, as done in Fig. \ref{fig2}. 
The $\theta$-dependence of $H_{\rm c2}(\theta)/H_{\rm c2,c}$ is presented 
in Figs. \ref{fig11}(b) and \ref{fig11}(c), 
where the filled (empty) symbols correspond to 
the first (second) order transitions.
Our calculation shows that the first order transitions 
only occur for $\theta=89^\circ,90^\circ,91^\circ$,
beyond which all $H_{\rm c2}$ transitions become second order 
as displayed in Figs. \ref{fig11}(b) and \ref{fig11}(c).
This is consistent to experimental data~\cite{kittaka} presented 
in the figures. 
There, in ${\rm Sr_2RuO_4}$ 
the first order transitions occur within $90^\circ\pm2^\circ$. 
The first order transition near the $ab$ plane appears because 
the effective paramagnetic parameter $\tilde{\mu}(\theta)=\mu \Gamma(\theta)$ 
in Eq. (\ref{eq:a2}) exceeds 
the critical value $\mu_{\rm cr}\sim 1.7$ for the first order transition
only for the angles $89^\circ<\theta<91^\circ$.

As seen from Figs. \ref{fig11}(b) and \ref{fig11}(c), 
$H_{\rm c2}(\theta)/H_{\rm c2,c}$ values by the numerical calculations 
are well fit by a simple function~\cite{Machida} 
in Eq. (\ref{eq:a3}) with $\mu=0.04$, explained in Appendix. 
The values of $H_{\rm c2}(\theta)/H_{\rm c2,c}$  
are slightly under the experimental values. 
This can be easily remedied by changing the $\mu$ value.
Namely, instead of the present value $\mu=0.04$,  
the refined value $\mu=0.0293$ 
shows much better fitting to the experimental data by Eq. (\ref{eq:a3}), 
as shown in Figs. \ref{fig11}(b) and \ref{fig11}(c).

We also evaluate 
the temperature dependence of the ratio 
$\Gamma_{H_{\rm c2}}(T) \equiv H_{\rm c2,ab}(T)/H_{\rm c2,c}(T)$, 
and compare it with the experimental data~\cite{kittaka2} 
in Fig. \ref{fig11}(d).
Near $T=T_{\rm c}$, both in numerical and experimental data, 
$\Gamma_{H_{\rm c2}}(T)$ shows the large anisotropy ratio,  
tending to $\sim 60$, which is governed by the Fermi velocity anisotropy
ratio of the $\beta$ band $\Gamma_{\beta}=57$.  
Upon decreasing $T$, this ratio progressively becomes small because of the PPE.
This is captured by our numerical calculation, and the 
tending limit towards the lowest $T$ is 16. 
The $T$-dependence is well fitted by Eq. (\ref{eq:a3}) with 
Eq. (\ref{eq:a4}) for $\mu=0.04$, as shown in Fig. \ref{fig11}(d). 
In the experimental data which reduces 20 at low $T$~\cite{kittaka2} 
is fitted by Eq. (\ref{eq:a3}) with $\mu$=0.0293 as in the case of 
Figs. \ref{fig11}(b) and \ref{fig11}(c). 
We also note that the fitting line with $\Gamma=180$ 
largely deviates from the experimental data. 
A similar analysis on the $H_{\rm c2}$ anisotropy data~\cite{kittaka2} 
is performed by Choi~\cite{choi} to come to the same conclusion.
We point out again that in numerical calculation (blue square) 
the point in low $T$ range in Fig. \ref{fig11}(d) corresponds 
to the first order $H_{{\rm c2}, ab}$ transition, 
while at least above $T/T_{\rm c}>0.5$, $H_{{\rm c2}, ab}$ 
is of second order. 
According to the experiments~\cite{yonezawa,kittaka}, 
the first order line at $H_{\rm c2,ab}(T)$ extends to around 
$T/T_{\rm c}>0.4\sim0.5$.
The  accurate termination point  between the first and second order transitions
will be a future problem.

In summary of this section, 
in both estimations of $\Gamma_{\rm VL}(\theta)$ and 
$\Gamma_{H_{\rm c2}}(T)$, 
the intrinsic anisotropy $\Gamma$ of ${\rm Sr_2RuO_4}$ is 
identified as $\Gamma=60$. 
Since it corresponds to $\Gamma_\beta=57$ given by 
the $\beta$ band among the known three bands.   
The anisotropy $\Gamma=60$ indicates that 
the $\beta$ band is fully responsible for determining various observables. 
Thus this should be the major band,  
while the $\gamma$ band with $\Gamma_\gamma=174$ 
is not appropriate for the major band as seen from Fig. \ref{fig11}(d)
and must be the minority band and the $\alpha$ band plays a negligible
role because its DOS is 10\% of the total.
Those considerations partly justify the present single band model to
grasp the essential points. 

\section{Conclusion and unsolved problems}
\label{sec:conclusion}

The essential assumption in the present theory is that 
the PPE works in the spin part of the pairing function. 
The assumption for the orbital part as isotropic $s$-wave 
is not intrinsic in our calculations. 
As the pairing function inducing the PPE, 
in addition to the spin-singlet pairing, 
the spin-triplet pairing is also available 
if the $d$-vector has components in the $ab$ plane.

There exist several outstanding experiments to claim 
as the evidence for a spin-triplet 
chiral p-wave pairing realizing in Sr$_2$RuO$_4$.
Among them, 
only the Knight shift experiments~\cite{ishida,murakawa,ishida2,hayden} 
by NMR and polarized neutron scattering 
are treated as an evidence for that the spin part of the pairing function 
is the spin-triplet. 
There the Knight shift remains 
unchanged for any magnetic field orientations, 
any $H$, $T$ and any nuclear species available so far. 
These are against the present theory.  
However, the unchanged $T$-dependence for all field orientations is a mystery, 
since the Knight shift is expected to be decreased 
for some of field orientations 
even in the spin-triplet pairing. 

Other experiments than the Knight shift are evidences  
for the orbital part of the pairing. 
There, the time reversal symmetry breaking is suggested by 
muon spin rotation ($\mu$SR)~\cite{luke}, 
which shows the appearance of the spontaneous moments below $T_{\rm c}$. 
These varieties of the orbital part can be easily accommodated 
in the present theoretical framework.
As for the time reversal symmetry breaking,
the possible pairing can be an $s+{\rm i}d$ or $d+{\rm i}d$ 
in the spin-singlet pairing, or $p_x+{\rm i}p_y$ in the spin-triplet pairing. 
Even in the spin-triplet pairing, the PPE works 
if the $d$-vector has a component within the $ab$ plane. 
The essential characteristics of 
the mixed state magnetic properties investigated here 
remains intact even for those pairings. 
In the present paper we have taken the isotropic $s$-wave
just for computational convenience and for the illustrative purpose.

From the results that the PPE plays significant contributions 
on the superconducting properties for $\bar{\bf B} \parallel ab$, 
we recognize the importance of the further studies to determine 
the spin part of the pairing function. 
In ${\rm Sr_2RuO_4}$, due to the spin-orbit coupling of the electronic states, 
we have to consider the coupling of the spin and the orbital parts  
in the spin-triplet pairing function~\cite{yanase,ng,scaffidi,tsuchiizu}. 
This may be an essential factor to realize a spin-triplet superconductor 
where the $d$-vector has the component in the $ab$ plane. 

Several important experimental facts remain unanswered in the present theory:

\noindent
(1) 
As shown in Fig. \ref{fig5}(c), 
the transverse form factors as a function of $\theta$ 
under a fixed field in the SANS experiment~\cite{Rastovski} exhibit
a strong decay of its amplitudes 
long before reaching the expected $H_{\rm c2}$,
when $\theta$ decreases away from the $ab$ plane at $\theta=90^\circ$
in the lower fields. 
We anticipate the interplay of the multiband effect 
to explain it~\cite{nakai}. 

\noindent
(2) It is also obvious that the $\gamma(H)$ behavior shown 
in Fig. \ref{fig6}(d) is quite different from the observed $C/T$ 
at low $T$ especially at lower field regions~\cite{deguchi} 
where a $\sqrt H$-like sharp increase is observed. 
This  was interpreted as coming from the minor band ($\alpha+\beta$)
that is assumed to have a line node like gap structure 
simply because the DOS value attained 
in that field region seems to that for 
the $\alpha+\beta$ bands (43\% of the total DOS). 
The $\gamma$ band with 57\% of the total DOS is regarded as the major band. 

This interpretation is at odds with the present
theory in several points; If this is true, 
the ``major'' $\gamma$ band should be responsible for the high field region.
Since the Fermi velocity anisotropy $\Gamma_{\gamma}=174$ for the $\gamma$ band,
we would expect $\Gamma_{\rm VL}\sim 174$ for $\bar{\bf B}\parallel ab$.
There is no indication  for it 
in the available SANS data~\cite{Rastovski} and other~\cite{kittaka2}. 
Instead, the SANS data~\cite{Rastovski} $\Gamma_{\rm VL}\sim 60$ 
indicate that the high field phase should be the $\beta$ band.
We point out also the data~\cite{kittaka2} 
$H_{\rm c2,ab}(T)/H_{\rm c2,c}(T)$ shown in Fig. \ref{fig11}(d) which
is directly related to the Fermi velocity anisotropy 
in Eq. (\ref{eq:Gamma-xi}) near $T_{\rm c}$.
Thus  the $\beta$ band is also responsible for it~\cite{nakai}. 
There is no trace in the existing data to show 
that the  the $\gamma$ band plays a major role.
The present single band theory assumes the $\beta$ band as the major band,
neglecting the minor $\gamma$ band. 
We need to refine it by taking into account
both bands in addition to the remaining $\alpha$ band.
A multiband  theory  based on Eilenberger framework belongs to 
a future work~\cite{nakai}. 

In order to further advance the ${\rm Sr_2RuO_4}$ problem 
concerning its pairing symmetry and multiband nature, 
we propose here several decisive experiments:

\noindent
(A) The SANS experiments to observe the longitudinal component $F_z$ 
in the FLL form factors.
As already predicted in Figs. \ref{fig4} and \ref{fig5}(c) 
the magnitudes $F_{z (1,0)}$ and $F_{z (1,1)}$ near $H_{\rm c2}$
are within the observable range. We expect to see 
the enhanced $F_z$ behavior towards $H_{\rm c2}$,
a similar behavior already observed in ${\rm CeCoIn_5}$~\cite{white} 
which is known to be a typical superconductor with strong PPE.

\noindent
(B) To determine the detailed gap structure on 
the $\alpha$, $\beta$ and $\gamma$ bands
the field angle resolved specific heat experiment is decisive. 
The existing data~\cite{deguchi} at low fields are only down to 100 mK
which was concluded to have a $d_{xy}$ like nodal structure, judging
from the existing four fold oscillation pattern.
We expect that the sign of the four fold oscillation in $C/T$ might change 
because the $\gamma$ band responsible
for this oscillation at low fields and low temperatures
should have a $d_{x^2-y^2}$ like nodal structure. Such a sign change
of the oscillation has been observed in CeCoIn$_5$~\cite{an}.  
This $d_{x^2-y^2}$ like nodal structure on the $\gamma$ band
is fully consistent with other experiments.
In particular, the square vortex lattice pattern oriented to (110) direction
is observed for $\bar{\bf B}\parallel c$ 
in the SANS experiments~\cite{forgan2}.
Note in passing that the gap structures of $\beta$ and $\alpha$ bands 
are relatively isotropic~\cite{nakai}. 
This might be consistent with the $c$-axis tunneling data~\cite{suderow} 
that probes selectively
the $\beta$ band with the least Fermi velocity 
along this direction and shows a full gap.

\noindent
(C) Finally according to the present theory with PPE, 
${\rm Sr_2RuO_4}$ is quite likely to exhibit 
the Fulde-Ferrell-Larkin-Ovchinnikov (FFLO) phase 
in low $T$ and high $H$ regions just below $H_{\rm c2}$.
One of the best way to detect it is to measure $T_1^{-1}$ by NMR,
which is enhanced in the FFLO phase. This method is successfully
applied to the organic superconductor 
$\kappa$-(BEDT-TTF)$_2$Cu(NCS)$_2$~\cite{vesna}. 

In summary,  we investigate the mixed state properties of a 
uniaxially anisotropic superconductor with the Pauli paramagnetic 
effect on the basis of microscopic Eilenberger theory in the clean limit,
assuming a single band model.
By these studies, 
we discussed the field orientation dependence near $\bar{\bf B} \parallel ab$, 
and tried to explain curious behaviors in ${\rm Sr_2RuO_4}$,  
focusing on contributions by the Pauli-paramagnetic effect. 
In the study of the longitudinal and transverse components of 
the flux line lattice form factors, 
effects of $H_{{\rm c2},ab}$ suppression due to the paramagnetic pair breaking 
is important to quantitatively explain the intensity of 
the spin-flip SANS experiment observed in ${\rm Sr_2RuO_4}$~\cite{Rastovski}. 
In the magnetization curve and field-dependent Sommerfeld coefficient, 
the jumps at the first order $H_{{\rm c2},ab}$ 
transition mainly come from the contribution 
of the paramagnetic susceptibility. 
From the study on the field orientation dependence of torque curves 
and $H_{\rm c2}(\theta)$, 
the intrinsic anisotropy ratio between $c$ and $ab$ directions is 
$\Gamma \sim 60$, suggesting the main contribution of the $\beta$-band 
in the superconductivity of ${\rm Sr_2RuO_4}$. 
These consistent behaviors between experimental observation and 
the theoretical calculation indicate that 
the Pauli paramagnetic effect plays important roles 
to understand curious behaviors 
at high fields for $\bar{\bf B}\parallel ab$ in ${\rm Sr_2RuO_4}$.  
This suggests that the pairing symmetry is either a spin-singlet or
a spin-triplet pairing with the $d$-vector components in the $ab$ plane.
We expect further experimental and theoretical future studies 
to confirm the mechanism of high field behaviors 
for $\bar{\bf B}\parallel ab$ in ${\rm Sr_2RuO_4}$.  

\begin{acknowledgments}
We thank T. Sakakibara, S. Kittaka, A. Kasahara, S. Yonezawa, 
Y. Maeno, M. R. Eskildsen, K. Ishida and Y. Matsuda 
for fruitful discussions and support from the experimental side.
\end{acknowledgments}

\appendix 
\section{}
\label{Appendix}

According to Ref. \onlinecite{Machida},  
an analytic expression for the $\mu$-dependence of the upper critical field 
$H_{\rm c2}(\mu)$ was derived 
by the fitting to the numerical solutions of Eilenberger equation 
under PPE at low temperatures. 
The expression for $H_{\rm c2}(\theta,\mu)$ is given by 
\begin{eqnarray} && 
\frac{H_{\rm c2}(\theta,\mu)}{H_{\rm c2}^{\rm orb}(\theta)} 
=\frac{1}{\sqrt{1+2.4 \tilde{\mu}(\theta)^2}}
\label{eq:a1}
\end{eqnarray} 
for each field orientation $\theta$ in a
uniaxial superconductor.  
There 
$H_{\rm c2}^{\rm orb}(\theta)=H_{\rm c2}(\theta,\mu=0)
=H_{{\rm c2},c}^{\rm orb}\Gamma(\theta)$ is 
the orbital limited upper critical field without PPE.
The effective paramagnetic parameter $\tilde{\mu}(\theta)$ at each $\theta$ 
depends on anisotropy $\Gamma(\theta)$ in Eq. (\ref{eq:Gamma}) as  
\begin{eqnarray} && 
\tilde{\mu}(\theta)
=\mu \frac{H_{\rm c2}^{\rm orb}(\theta)}{H_{{\rm c2},c}^{\rm orb}}
=\mu \Gamma(\theta) ,
\label{eq:a2}
\end{eqnarray} 
since $\tilde{\mu}(\theta) \propto H_{\rm c2}^{\rm orb}(\theta)/H_p$.  
The Pauli paramagnetic critical field $H_p$ 
is proportional to the pair potential $\Delta$ at a zero field. 
By combining Eqs. (\ref{eq:a1}) and (\ref{eq:a2}), we obtain 
\begin{eqnarray} && 
\frac{H_{\rm c2}(\theta,\mu)}{H_{\rm c2, c}^{\rm orb}}
= \frac{1}{\sqrt{\cos^2\theta+\Gamma^{-2}\sin^2\theta +2.4 \mu^2}}. 
\label{eq:a3}
\end{eqnarray} 
This gives the $\theta$-dependent $H_{\rm c2}(\theta)$ 
in Figs. \ref{fig11}(b) and \ref{fig11}(c).
When we evaluate the $T$-dependence of 
$H_{\rm c2}(\theta,\mu)/H_{\rm c2, c}^{\rm orb}$ in Eq. (\ref{eq:a3}), 
$\mu=\mu(T=0)$ is modified to
the $T$-dependent paramagnetic parameter $\mu(T)$ 
given by 
\begin{eqnarray}  
\mu(T) 
= \mu \frac{\Delta(0)}{\Delta(T)} 
      \frac{H_{{\rm c2},c}^{\rm orb}(T)}{H_{{\rm c2},c}^{\rm orb}(0)}.   
\label{eq:a4}
\end{eqnarray}
The $T$-dependent orbital limited $H_{{\rm c2},c}^{\rm orb}(T)$ 
is given by the Werthamer-Helfand-Hohenberg (WHH) formula~\cite{WHH}
or the solution of the Eilenberger equation in the clean limit. 
The $T$-dependent order parameter $\Delta(T)$ follows the BCS form.
The $T$-dependence of  
$H_{\rm c2}$ anisotropy $H_{\rm c2, ab}(T)/H_{\rm c2, c}(T)$ 
when $\theta=90^\circ$ is 
obtained by Eqs. (\ref{eq:a3}) and ({\ref{eq:a4}),  
which is displayed in Fig. \ref{fig11}(d).


\end{document}